\documentclass[twocolumn,preprintnumbers,amsmath,amssymb]{revtex4}

\usepackage{graphicx}
\usepackage{dcolumn}
\usepackage{bm}

\begin{document} 
  

\title{$K$-Scaffold subgraphs of Complex networks}

\author{Bernat Corominas-Murtra$^1$, Sergi Valverde$^1$,
Carlos Rodr\'iguez-Caso$^1$ Ricard V. Sol\'e$^{1,2}$}

\affiliation{ $^1$ ICREA-Complex Systems
Lab,  Universitat Pompeu  Fabra,  Dr.  Aiguader  80, 08003  Barcelona,
Spain\\  $^2$ Santa  Fe Institute,  1399  Hyde Park  Road, New  Mexico
87501, USA }

\begin{abstract} Complex networks with  high numbers of nodes or links
are often  difficult to analyse. However, not  all elements contribute
equally to their structural patterns.  A small number of elements (the
hubs)  seem to  play a  particularly relevant  role in  organizing the
overall  structure around them.  But other  parts of  the architecture
(such  as hub-hub  connecting elements)  are also  important.  In this
letter  we  present  a new  type  of  substructure,  to be  named  the
$K$-scaffold  subgraph,  able to  capture  all  the essential  network
components.  Their  key features,  including  the  so called  critical
scaffold graph, are analytically derived.
  \end{abstract}
  
\maketitle

\noindent
 {\em      Introduction}.        Networks      pervade      complexity
\cite{ReviewBA,Newmann,  DorogovtsevReview, Latora}. How  networks are
organized at  different scales  is one of  the main topics  of complex
network  research \cite{Vicsek,  Dorogovstevkcore1,  Itzkovitz, Aalon,
Toroczkai,  Skeletons}.  Some  approaches are  based on  the  study of
given  subgraphs,   from  the  smaller   network  motifs  \cite{Aalon,
Itzkovitz,    ValverdeMotifs}     to    $K$-cores    \cite{Vespingani,
Dorogovstevkcore1},   spanning  trees  \cite{Skeletons}   or  gradient
subgraphs   obtained   form  a   given   internal  system's   dynamics
\cite{Toroczkai}.

\noindent
One of the most studied  subgraphs is the so-called $K$-core, formally
defined  by Bollob\`as in  \cite{Bollobas}.  The  $K$-core of  a graph
${\cal  G}$,  ${\cal  C}_k(\cal  G)$  is the  largest  subgraph  whose
vertices have,  at least,  degree $k \geq  K$.  The behaviour  of such
subgraph,  and its  percolation  properties have  been widely  studied
\cite{Bollobas,Dorogovstevkcore1,  CoresFernholz, MolloyReed, Molloy}.
$K$-cores  display interesting features  with several  implications in
the study of real networks,  both at the theoretical and applied level
\cite{Dorogovstevkcore1,Vespingani, Almaas, CoresFernholz}.

\noindent
Hubs  are  the  center  of   attention  of  the  $K$-core. They are
responsible for  the efficient  communication among network  units and
their   failure   or    removal   can   have   dramatic   consequences
\cite{Barabasi}.   But other  graph  components are  also relevant  to
understand network  behavior.  In  particular, hubs are  often related
through other elements exhibiting low connectivity, the so-called {\em
conectors}.  Despite its relevance,  the $K$-core fails in finding the
hub-connector  structure.    This  pattern  is   essential  in  highly
dissassortative or  modular networks,  where hub-hub conectors  play a
crucial  role \cite{Newmann}.   In such  networks,  robustness against
failures  is strongly  tied  to  hubs, but  also  to the  hub-conector
structure.   Moreover,  conectors  can  display  high  {\em  betweness
centrality} \cite{Newmann}  despite their low  conectivity, reinforcing
the role of this kind of nodes in non-local organization of the global
topology and dynamics.

\noindent  To  overcome these  limitations,  we  introduce a  subgraph
definition  which  captures  the  previous traits.   Specifically,  we
consider a subgraph  that includes the most connected  nodes and their
connectors, if any.   In doing so, we want to  explore wether there is
some fundamental  hub-connector subgraph and  its relevant properties.
Such  a  graph,  the  so-called $K$-scaffold  subgraph,  was  recently
introduced  (in qualitative  terms) within  the context  of  the human
proteome  \cite{Shals}.   This  network  included  only  transcription
factors,  i.   e.   proteins  linking  to DNA  and  thus  involved  in
regulating  gene expression  (fig 1(a)).   Specifically, it  was shown
that  an  appropiate choice  of  relevant  hubs  and their  connectors
allowed to  define a functionally meaningful  subgraph.  Such subgraph
contained  a  large number  of  cancer-related  proteins around  which
well-defined modules were organized as evolutionarily and functionally
related subsets. Here,  we define this subgraph in  a rigorous way. We
analitically characterize  its properties and  degree distributions as
well  as the presence  of a  special class  of minimal  scaffold graph
based on a critical percolation threshold.
\begin{figure*} 
\includegraphics[width=13 cm]{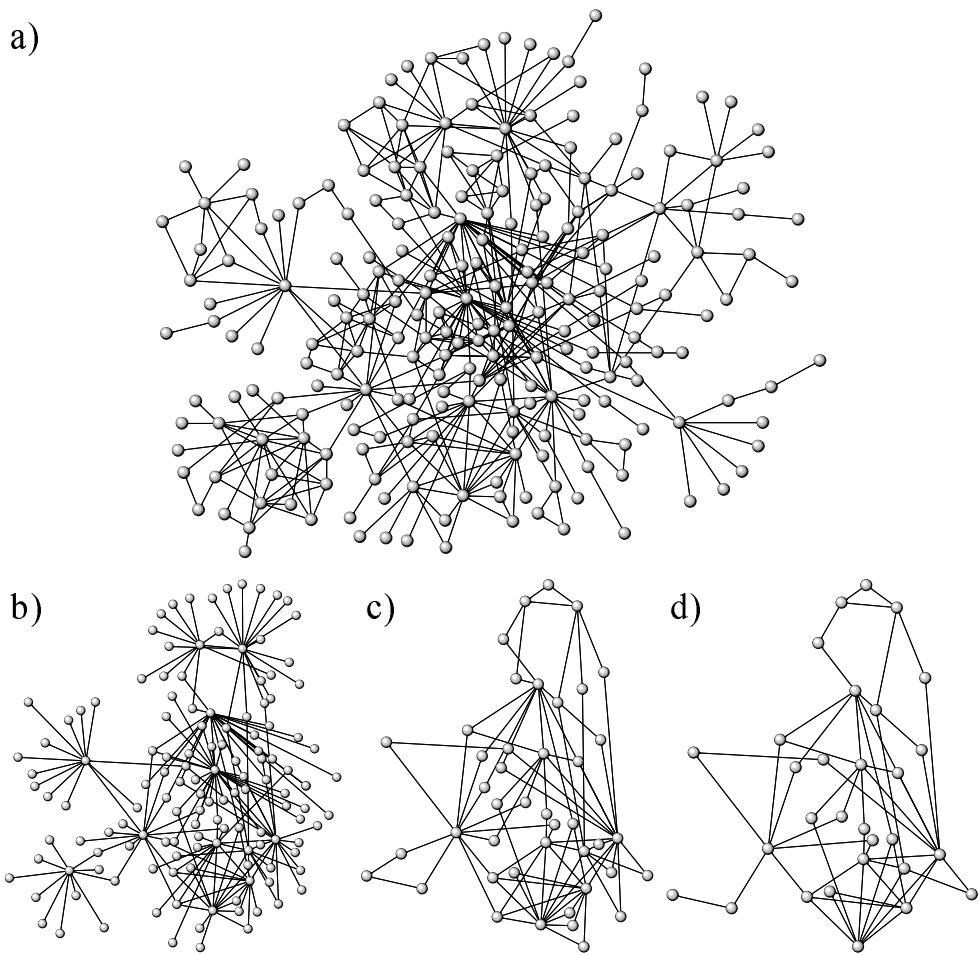}
\caption{(a)  The  Human   Transcription  Factor  interaction  Network
(HTFN).   (b)  its 11-Scaffold  subgraph.  (c)  The naked  11-scaffold
subgraph and (d) the  naked and renormalized 11-scaffold subgraph. The
$K$-scaffold subgraph  displays a fundamental  hub-connector structure
that organizes  the general topology  of the whole system.   Data from
\cite{Shals}.}
\label{TFScaffold}
\end{figure*}

$\\$
\noindent
{\em  $K$-Scaffold subgraphs}. Let  us consider  a graph  ${\cal G}(V,
\Gamma)$, where $V$ is the set  of nodes and $\Gamma$ the set of edges
connecting  them.   The  $K$-Scaffold  subgraph $S_K({\cal  G})$  will
consider the  degree of  nodes $k(e_i),\;e_i \in  E$ but it  will take
into account  correlations: Specifically, if we choose  a node $e_i\in
V$, it will belong to $S_K({\cal G})$ if and only if $(1)$ $K \le k_i$
or $(2)$ $e_i$ is connected to $e_j  \in V$, and $e_j$ is such that $K
\le  k_j$.   Thus, given  a  graph  ${\cal  G}(V, \Gamma)$,  with  its
adjacency  matrix $a_{ij}$,  the $K$-scaffold  of ${\cal  G}$  will be
defined as:
\begin{equation}
  S_{ij}=\left\{
  \begin{array}{ll}
    a_{ij} \;{\rm iff}\; \left( K \le k_i\;\vee \; K \le k_j \right)\\
    0\;{\rm otherwise}
  \end{array}
  \right.
\end{equation}
\noindent
An example of  such $K$-scaffold subgraph is shown  in (fig.(1b)). This
allows us to define, from a  given graph ${\cal G}$ a nested hierarchy
of subgraphs $S_K({\cal G})$ such that:
\begin{equation}
...S_{K+1}({\cal G}) \subseteq S_K({\cal G}) \subseteq S_{K-1}({\cal G})...
\end{equation}
\noindent  To clarify  which  elements are  really  relevant, we  also
define  a {\em  naked} $K$-scaffold  subgraph.  From  the $K$-scaffold
subgraph, $S_K({\cal  G})$, the $naked$  $K$-scaffold, $\gamma_K({\cal
G})$,  is obtained by  removing all  nodes having  a single  link (the
``hair'' of the graph)(fig.(1c)).  Thus,  from $S_{ij}$, it is easy to
compute the adjacency matrix of the {\em naked} $K$-scaffold subgraph,
namely:
\begin{equation}
\gamma_{ij}=S_K(a_{ij})(1-\delta_{k_i,1})(1-\delta_{k_j,1})
\end{equation}
\noindent
Additionally,  if two  or more  connectors have  identical  pattern of
conectivity  in  $\gamma_K({\cal G})$  (i.e.,  they  are connected  to
exactly the same hubs, understanding hubs as nodes with $k \ge K$), we
renormalize these sets of connectors  by replacing each of them with a
single  node.  In  this way,  the renormalized  $K$-scaffold subgraph,
$\overline{\gamma_K}({\cal G})$,  keeps the relevant  elements without
redundancies (fig.(1d)).

$\\$
\noindent  {\em  Statistical Properties.}   Here  we  derive the  main
statistical features  of the $K$-scaffold subgraph  from an arbitrary,
uncorrelated network ${\cal  G}$ .  First, we compute  the fraction of
nodes  in   ${\cal  G}$  belonging  to  $S_K({\cal   G})$,  i.e.,  the
probability  for a  random choosen  node of  ${\cal G}$  to  belong to
$S_K({\cal G})$. If we define
\begin{equation}
q_{<K} =\sum_{k<K}\frac{kP(k)}{\langle k \rangle_{\cal G}}
\end{equation}
\noindent
We can define $f$ as:
\begin{equation}
f=1-\sum_{k<K}P(k)\left(q_{<K}\right)^k
\label{FracScaf}
\end{equation}
\noindent  Where  we  have  to   read  the  second  term  of  equation
(\ref{FracScaf}) as  {\em the  probability to find  a node  with $k<K$
such that all of its $k$ first neighbours have $k'<K$}.

\noindent We could consider  or not links connecting connectors, i.e.,
nodes with  $k<K$ but connected to  nodes with $k\geq K$.  In order to
simplify  algorithmic  procedures,  it  is reasonable  to  avoid  such
connector-connector  links.  If  we avoid  these class  of  links, the
probability for a randomly choosen  link to belong to the $K$-scaffold
is, simply:
\begin{equation}
h=1-(q_{<K})^2
\end{equation}
\noindent However,  for mathematical  consistency, it is  necessary to
take into account such kind of links. If we do so, the probability for
a randomly choosen link to belong to the $K$-scaffold is:
\begin{equation}
h=1-\left(\sum_{k<K}q(k)(q_{k<k})^{k-1}\right)^2
\end{equation}
\noindent
To complete  our characterization, we find the  degree distribution of
$S_K({\cal G})$.   To do  the job, we  need to define  the probability
that a  node whith degree  less than $K$  is connected to  exactly $k$
nodes whose connectivity is equal or higher than $K$, namely:
\begin{equation}
g(k,K)=\sum_{k<i<K}P(i)(1-q_{<K})^k(q_{<K})^{i-k}
\end{equation}
\noindent
The probability  distribution for $k$'s above $K$  in the $K$-scaffold
is  the same probability  distribution of  the substrate  graph ${\cal
G}$, multiplied by a normalization factor $f$. To see this, we can see
that, from the  definition of $g(k,K)$ we have:
\begin{equation}
\sum_{k<K}g(k,K)+\sum_{k \ge K}^{\infty}P(k)=
\frac{|N|_{S_k({\cal G})}}{|N|_{\cal G}}=f
\end{equation}
\noindent
Thus, the normalized degree distribution of $S_K({\cal G})$ will be:
\begin{equation}
P_{S_K}(k)=\left\{
\begin{array}{ll}
  g(k,K)/f,\; {\rm iff} \;  k<K\\ 
  P(k)/f, \; {\rm otherwise}
\end{array}
\right.
\end{equation}

$\\$
\noindent   {\em  Minimal   $K$-Scaffold  subgraphs}.    The  previous
definitions  refer to  $K$-dependent subgraphs,  being  $K$ arbitrary.
But we  can ask if some  specially relevant $K$ value  is involved. In
otherwords, since  larger $K$ values support  smaller scaffold graphs,
we might  ask what  is the limit  in this  process and what  is keeped
before the  network is  fragmented or too  small. The question  we are
addressing concerns the existence  of a characteristic scale. The {\em
minimal}  scaffold  subgraph   will  label  the  minimal  substructure
capturing the fundamental hub-connector architecture of the net, if it
exists.  This subcritical subgraph will be located immediatly above of
the  percolation   threshold  of  ${\cal  G}$,   considering  how  the
$K$-scaffold performs node deletion.
\begin{figure*}
\includegraphics[width=17.5 cm]{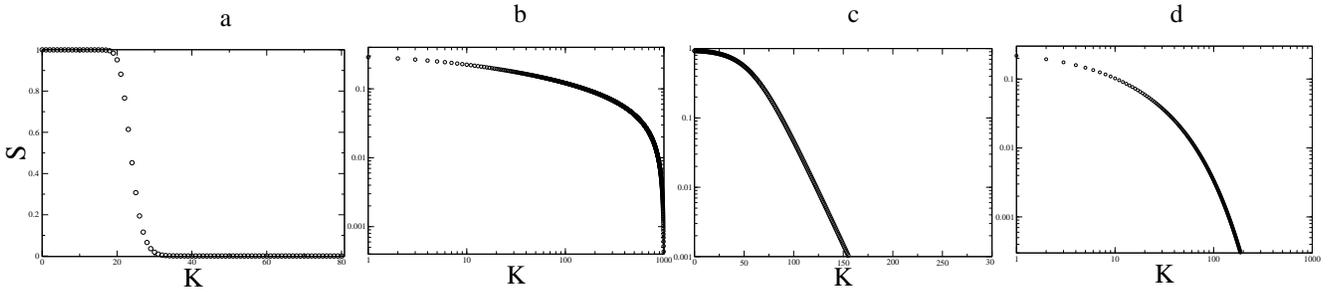}
\caption
{Numerical  simulations of the  relative size  of the  giant component
${\cal   S}_K$  of   the  $K$-scaffold   subgraphs  of   model  degree
distributions  against  $K$.   ${\cal  S}_K$  is  computed  as  ${\cal
S}_K=F_0(0)-F_0(v)$, where $v$ is  the numerical approach of the first
non-trivial,  physically   relevant  solution  of   the  the  equation
$v=1-F_1(1)+F_1(v)$ (See  text).  The  plots display ${\cal  S}_K$ for
(a)  Erd\"os -R\'enyi  graph,  whith $\langle  k  \rangle=15$.  (b)  A
Scale-free network  with $P(k)\propto k^{-\alpha}$,  $\alpha=2.3$.  No
$K_c$  can be  properly identified  (see text).   The presence  of the
cut-off can be  due to the finite size effects  of the simulation; (c)
Exponential net with $P(k)\propto  e^{-k/{\cal K}}$, ${\cal K}=13$ (d)
Power   Law   with  exponential   cut   off   net  with   $P(k)\propto
k^{-\alpha}e^{-k/{\cal K}}$, for $\alpha=2.5$ and ${\cal K}=52$. }
\label{Kcs}
\end{figure*}
\noindent
Following  the configuration  model \cite{Besesky},  we  consider very
large   random   graphs   with   an  arbitrary   degree   distribution
\cite{Random}.  Thus,  given a graph ${\cal G}$  we compute $S_K({\cal
G})$  by increasing  $K$ until  it breaks  down into  many unconnected
components. The probability  for a node to belong  to the $K$-scaffold
subgraph will be a function of  its connectivity $k$. We will refer to
this function as $f_k$:
\begin{equation}
f_k=\left\{
\begin{array}{ll}
1,\; {\rm iff}\; k \ge K \\
1-\left(q_{<K}\right)^k, \;{\rm otherwise}
\end{array}
\right.
\end{equation}
\noindent
Clearly,   a    node   with    degree   $k<K$   has    a   probability
$1-\left(q_{<K}\right)^k$ to  be connected with  at least to  one node
with degree higher than $K$.
\noindent
Now  we  define  the  generating  functions  \cite{Dorogovstevkcore1,
Random,  Moore,  Callaway}, taking  in  account  that  to compute  the
$K$-scaffold implies deleting  a given fraction nodes \cite{Callaway},
namely:
\begin{equation}
F_0(x)=\sum_{k}^{\infty}P(k)f_k x^k
\end{equation}
\noindent
Let us  define $F_1(x)$ as:
\begin{equation}
F_1(x)=\frac{1}{\langle k \rangle_{\cal G}}\sum_k^{\infty}kP(k)f_kx^{k-1}=\frac{1}{\langle k \rangle_{\cal G}}
\frac{dF_0(x)}{dx}
\end{equation}
\noindent  Using  previous  theoretical results  \cite{Random,  Moore,
Callaway},  we compute  the generating  functions for  the probability
distribution  of component sizes  other than  the giant  component, if
any.   $H_1(x)$ will  be defined  as the  generating function  for the
probability that  one end  of a randomly  choosen edge on  the network
${\cal  G}$ -when  computing its  $K$-scaffold- leads  to  a connected
component of  a given number  of nodes. This includes  the probability
that such component  will contain zero nodes, because  of the deletion
of  nodes  of ${\cal  G}$  when  computing  the $K$-scaffold.   As  we
discussed    above,   this    will   happen    with    a   probability
$1-f=1-F_1(1)$. The end of the edge can be occupied by a node with $k$
outgoing edges,  distributed along $F_1(x)$  \cite{Callaway}. Thus, it
leads  us  to   the  self-consistency  equation  \cite{Random,  Moore,
Callaway} -for a  clear and detailed derivation of  these results, see
\cite{Moore, NewmanCondMat}:
\begin{equation}
H_1(x)=1-F_1(1)+xF_1(H_1(x))
\label{H1}
\end{equation}
\noindent
And the generating  function for the size of the  component to which a
random choosen node belongs will be \cite{Random,  Moore,  Callaway}:
\begin{equation}
H_0(x)=1-F_0(1)+xF_0(H_1(x))
\label{H0}
\end{equation}
\noindent The  size of the giant  component ${\cal S}_K$  that we will
further  identify  with  the  $K$-scaffold  subgraph  will  be  
\begin{equation}
{\cal S}_K=F_0(1)-F_0(v)
\end{equation}
\noindent
where $v$  is the  first non  trivial, physically
relevant solution of 
\begin{equation}
v=1-F_1(1)+F_1(v)
\end{equation}
$\\$

\noindent We  can now look  for a singularity  in the average  size of
components.   Immediately above of  this point  we define  the minimal
$S_K({\cal     G})$.     Knowing     that    
\begin{equation}
\langle     s    \rangle =\left.\frac{dH_0(x)}{dx}\right|_{x=1}
\end{equation}
 \noindent we find, after  some   algebra  \cite{Random,   Moore,
Callaway}:
\begin{equation}
\langle s \rangle=F_0(1)+\left.\frac{d F_0(x)}{dx}\right|_{x=1}\times \frac{F_1(1)}{1-\left.\frac{dF_1(x)}{dx}\right|_{x=1}}
\end{equation}
\noindent  With a  singularity  when $[dF_1(x)/dx]_{x=1}=1$. Now using:
\begin{equation}
\left.\frac{d F_1(x)}{dx}\right|_{x=1}= \frac{1}{\langle k \rangle_{\cal G}}\sum_k^{\infty}k(k-1)f_kP(k)
\label{Criticallity}
\end{equation}
\noindent
we can derive the percolation  condition for a $K$-scaffold graph from
a  given substrate  graph  ${\cal G}(V,  \Gamma)$  with given  average
conectivity  $\langle  k  \rangle_{\cal  G}$ and  degree  distribution
$P(k)$.   We  compute  such  condition  taking in  account  the  above
critical  condition (\ref{Criticallity}).   (Recall  that computations
are  performed considering  the successive  pruning of  ${\cal  G}$ by
increasing $K$). Knowing that:
\begin{equation}
\sum_{k}^{\infty}\frac{kP(k)}{\langle k \rangle_{\cal G}}=1
\end{equation}
\noindent
and since equation (\ref{Criticallity}) is equivalent to:
\begin{equation}
\sum_k^{\infty}k(kf_k-f_k-1)P(k)=0
\end{equation}
\noindent
the percolation condition for a $K$-scaffold, $S_K$, is:
\begin{equation}
\sum_k k(kf_k-f_k-1)P(k)>0
\label{Spercolation}
\end{equation}
\noindent
We can easily see such a  condition as the extension of the Molloy and
Reed criteria \cite{Molloyreed} for the $K$-scaffold, $S_K({\cal G})$:
\begin{equation}
\sum_k^{\infty}k(k-2)P(k)> \sum_{k<K}k(k-1)q_{<K}^kP(k)
\label{Spercolation1}
\end{equation}
\noindent   Note   that   the   right-hand  side   of   the   equation
(\ref{Spercolation1})  is always  finite, whereas  the  left-hand side
could  not   be  finite.  Indeed,  note  that   $q_{<K}\le  1$,  thus,
\begin{eqnarray}
(q_{<K})^k\le   1
\end{eqnarray}
\noindent
Futhermore, $P(k)\le 1$.  Thus, the sum 
\begin{equation}
\sum_{k<K}k(k-1)q_{<K}^kP(k)<\infty
\end{equation}
\noindent
is always finite  provided that $k$ is bounded.   It is straighforward
that:
\begin{equation}   
 \sum_k^{\infty}k(k-2)P(k)=\langle    k^2   \rangle-2\langle k \rangle
\label{Spercolation2}
\end{equation}

\noindent Thus, a finite $K$ for the critical scaffold will exist {\em
if and  only if} the degree  distribution of the network  has a finite
second moment  $\langle k^2  \rangle$.  An Erd\"os-R\'enyi  graph, for
example, will display a  critical $K$-scaffold, provided that $\langle
k^2  \rangle_{ER}=\langle  k  \rangle^2$.   But  for  arbitrary  large
scale-free networks  with realistic exponents  ($2<\alpha<3$), we find
that  there is  not such  minimal $K$-scaffold.   This is  due  to the
divergence of the second moment. Thus, condition (\ref{Spercolation1})
always  applies  for  all  $K$'s.   Implications  should  be  studied,
provided that  it implies that  the hub-conector structure  appears at
all scales.

\noindent  Numerical simulations  (Fig. (\ref{Kcs}))  show that  if we
introduce a cut-off in  the degree distribution, a characteristic scale
$K_c$  is  present.  In  E-R  graphs,  the  size of  the  $K$-scaffold
displays  an abrupt decay  beyond $\langle  k \rangle$.   Finally, the
size of ${\cal S}_K({\cal G})$  displays a critical $K$ in exponential
networks.  Note that by its definition, if $S_K({\cal G})$ percolates,
also  do   the  corresponding  naked   and  renormalized  counterparts
(${\gamma_K}({\cal G})$ and $\overline{\gamma_K}({\cal G})$).

$\\$
\noindent  {\em  Discussion}.  $K$-Scaffold  subgraphs can  be  easily
measured on any arbitrary network and can be useful to detect both key
elements and the topological components that glue them. If topological
organization is linked with functionality, particularly in relation to
hubs, the scaffold of a complex  network should be able to capture the
relevant  subsystem.   For  the  human  transcription  factor  network
\cite{Shals} it  was found  that for $K=11$,  a small set  of proteins
having  relevant   cellular  functions  (including   oncogenes,  tumor
supressor genes and the  TATA-binding protein) was obtained, being all
of them  related through intermediate connector proteins.   This is in
agreement    with   the    dissasortative   character    of   celullar
networks. Since  each hub  was associated to  a group  of functionally
related TFs, the connectors  were actually relating different parts of
the protein machinery. Other real  systems have also been analysed and
provided  further  confirmation  of  the  relevance  of  the  scaffold
approach (Corominas Murtra et  al, in preparation). Further extensions
and properties of  this subgraph, togheter with te  analysis of finite
size effects associated to real  systems will be presented elsewhere.


 

\begin{acknowledgments} 
The authors thank  the members of the Complex  Systems.  This work has
been supported  by grants  FIS2004-0542, IST-FET ECAGENTS,  project of
the European Community founded under EU R\&D contract 01194, by the EU
within the 6th Framework Program under contract 001907 (DELIS), by NIH
113004, FIS2004-05422 and by the Santa Fe Institute.
\end{acknowledgments}

\end{document}